\documentclass[11pt]{article}
\usepackage[margin=1in]{geometry}
\usepackage{graphicx}
\usepackage{amssymb}
\usepackage{amsmath}
\usepackage{booktabs}
\usepackage{amsthm}
\usepackage[table]{xcolor}
\usepackage{stmaryrd}
\usepackage{mathabx}
\usepackage{mathrsfs}
\usepackage{latexsym}
\usepackage{rotating}
\usepackage{enumitem}
\usepackage[ruled,vlined,commentsnumbered,titlenotnumbered]{algorithm2e}
\usepackage{algorithmic}
\usepackage{url}
\usepackage{comment}
\usepackage{cite}

\excludecomment{comment}
\numberwithin{figure}{section}

\newtheorem{theorem}{Theorem}[section]
\newtheorem{lemma}[theorem]{Lemma}

\theoremstyle{definition}

\newcommand{\ignore}[1]{}

\newcommand{\ceil}[1]{\lceil #1 \rceil}

\newcommand{\lo}{\operatorname{lo}}
\newcommand{\hi}{\operatorname{hi}}
\newcommand{\contour}{\mbox{\sc contour}}
\newcommand{\row}{\operatorname{r}}
\newcommand{\col}{\operatorname{c}}

\newcommand{\ThreeSUM}{\textsf{3SUM}}
\newcommand{\TwoSUM}{\textsf{2SUM}}
\newcommand{\SUM}{\textsf{SUM}}

\newcommand{\LDT}{\textsf{LDT}}

\def\A{{\cal A}}
\def\reals{{\mathbb R}}

\newcommand{\Patrascu}{P\v{a}tra\c{s}cu}
\newcommand{\GP}{Gr\o nlund and Pettie~\cite{GP14}}

\title{Improved Bounds for 3SUM, $k$-SUM, and Linear Degeneracy\thanks{
		Work on this paper has been supported 
		by Grant 892/13 from the Israel Science Foundation, 
		by Grant 2012/229 from the U.S.-Israeli Binational Science Foundation,
		by the Israeli Centers of Research Excellence (I-CORE)
		program (Center No.~4/11),
		and by the Hermann Minkowski--MINERVA Center for Geometry at Tel Aviv
		University.}}

\author{Omer Gold\thanks{School of Computer Science, Tel Aviv University, Tel Aviv 69978,
		Israel; {\tt omergold@post.tau.ac.il}} 
	\and Micha Sharir\thanks{School of Computer Science, Tel Aviv University, Tel Aviv 69978,
		Israel; {\tt michas@post.tau.ac.il}}} 

\begin{document}
\sloppy

\begin{titlepage}
	
	\maketitle

\begin{abstract}
Given a set of $n$ real numbers, the \ThreeSUM{} problem is to decide whether there are three of them that sum to zero.
Until a recent breakthrough by Gr\o nlund and Pettie [FOCS'14],
a simple $\Theta(n^2)$-time deterministic algorithm for this problem was conjectured to be optimal.
Over the years many algorithmic problems have been shown to be reducible from the \ThreeSUM{} problem or its variants, including the more generalized forms of the problem,
such as $k$-\SUM{} and $k$-variate linear degeneracy testing ($k$-\LDT{}). 
The conjectured hardness of these problems have become extremely popular for basing conditional lower bounds for numerous algorithmic problems in P.

In this paper, we show that the randomized $4$-linear decision tree complexity\footnote{An $r$-linear decision tree is one in which each branching is based on a sign test of a linear expression with at most $r$ terms. The complexity of the tree is its depth.} of \ThreeSUM{} is $O(n^{3/2})$,
and that the randomized $(2k-2)$-linear decision tree complexity of $k$-\SUM{} and $k$-\LDT{}
is $O(n^{k/2})$, for any odd $k\ge 3$.
These bounds improve (albeit randomized) the corresponding $O(n^{3/2}\sqrt{\log n})$ and $O(n^{k/2}\sqrt{\log n})$
decision tree bounds obtained by Gr\o nlund and Pettie.
Our technique includes 
a specialized randomized variant of fractional cascading data structure. 
Additionally, we give another deterministic algorithm for \ThreeSUM{} that runs in $O(n^2 \log\log n / \log n )$ time. 
The latter bound matches a recent independent bound by Freund [Algorithmica 2017], but our algorithm is somewhat simpler, 
due to a better use of the word-RAM model.
\end{abstract}

\end{titlepage}

\section{Introduction}

The general \ThreeSUM{} problem is formally defined as
\setdescription{leftmargin=1.5cm,labelindent=\parindent}
\begin{description}
	\item[{\bf \ThreeSUM:}]
	Given a finite set $A \subset \mathbb{R}$, determine whether there exist $a,b,c\in A$ such
	that $a+b+c=0$.
\end{description}
An equivalent variant is that the input consists of three finite sets $A,\, B,\, C \subset \mathbb{R}$ of the same size, and the goal is to determine whether there are elements
$a\in A,\, b\in B,\, c\in C$ such that $a+b+c=0$. When the sets $A, B, C$ are not of the same size, the problem is named unbalanced \ThreeSUM{}.

The \ThreeSUM{} problem and its variants are among the most fundamental problems in algorithm design.
Although the \ThreeSUM{} problem itself does not seem to have many compelling practical implications,
it has been of wide interest due to numerous problems that can be reduced from it.
The notion of \ThreeSUM-Hardness is often used to describe such problems, namely, problems that are at least as hard as \ThreeSUM{}.
Thus, lower bounds on \ThreeSUM{} imply lower bounds on dozens of other problems.
Among them are fundamental problems in computational geometry~\cite{GajentaanO95,AichholzerADHRU12,BarequetH01,SossEO03}, dynamic graph algorithms~\cite{Patrascu10,AbboudW14, KopelowitzPP14},
triangle enumeration~\cite{AW15, KopelowitzPP14}, and pattern matching~\cite{AbboudWW14,AmirCLL14,ButmanCCJLPPS13, KopelowitzPP14, AKLPPS15}. 


In the last decades, starting with a study of Gajentaan and Overmars~\cite{GajentaanO95},
it was conjectured that any algorithm for \ThreeSUM{} requires $\Omega(n^2)$ time.
However, a recent breakthrough by \GP{} showed that \ThreeSUM{} can be solved in subquadratic time.
Specifically, they gave a {\em deterministic} algorithm that
runs in $O(n^2 (\log\log n / \log n)^{2/3})$  time, and a randomized algorithm that runs in $O(n^2 (\log\log n)^2 / \log n)$ expected time and with high probability.
Furthermore, they showed that there is a $4$-linear decision tree for \ThreeSUM{} with depth $O(n^{3/2}\sqrt{\log n})$
(i.e., the depth bounds the number of branching operations, each one is based on sign test of a linear expression with at most $4$ terms). 
These results raised serious doubts on the optimality of many algorithms for \ThreeSUM-Hard problems.
For example, the following problems are known to be \ThreeSUM{}-Hard.
(1) Given an $n$-point set in $\mathbb{R}^2$, determine whether it contains three collinear points (Gajentaan and Overmars~\cite{GajentaanO95}).
(2) Given $n$ triangles in $\mathbb{R}^2$, determine whether their union contains a hole, or compute the area of their union~\cite{GajentaanO95}.
(3) Given two $n$-point sets $X, Y \subset \mathbb{R}$, each of size $n$, determine whether all elements in $X+Y=\{x+y \;|\; x\in X,\; y\in Y \}$ are distinct (Barequet and Har-Peled~\cite{BarequetH01}).
(4) Given two $n$-edge convex polygons,  determine whether one can be placed inside the other via translation and rotation~\cite{BarequetH01}.

Problems 1 and 2 are solvable in $O(n^2)$ time (see~\cite{GajentaanO95}). Problems 3 and 4 are solvable in $O(n^2\log n)$ time (see~\cite{BarequetH01}).
In face of the new \ThreeSUM{} result of \GP{}, it is natural to ask whether these bounds are optimal. However, no better bounds are currently known (in spite of the improvement in~\cite{GP14}).
Problem 3 (or its stronger variant of sorting $X+Y$) has special importance, as it is used for basing the conditional lower bounds for the problems in~\cite{BarequetH01} and in~\cite{Barrera96}; these problems are therefore also classified as ``(Sorting $X+Y$)-Hard".
The decision tree complexity of Problem 3 was shown to be $O(n^2)$ by Fredman~\cite{Fredman76}.
It is a prominent long-standing open problem whether Problem 3 can be solved in $o(n^2\log n)$ time (see~\cite{OpenProblemsProject}).

In view of the results in~\cite{GP14}, the \ThreeSUM{} conjecture has been replaced by a 
relaxed, modern variant, asserting that \ThreeSUM{} cannot be solved in {\em strongly subquadratic time} (even in expectation), i.e., in $O(n^{2-\epsilon})$ time,
for any $\epsilon >0$.
This conjecture is widely accepted and believed by the computer science community, and so are its implications for deriving
lower bounds for other problems.
Abboud and Vassilevska-Williams~\cite{AW15} argue, based on the collective
computer science community efforts,
that lower bounds that are based on the relaxed \ThreeSUM{} conjecture
should be at least as believable as any other known conditional lower bounds for a problem in P.

The \ThreeSUM{} problem was also extensively studied in its generalized forms, $k$-\SUM{} and $k$-variate linear degeneracy testing ($k$-\LDT{}), formally defined as
\begin{description}
	\item[{\bf $k$-\LDT{} and $k$-\SUM:}]
	Given a $k$-variate linear function $\phi(x_1,\ldots,x_k) = \alpha_0 + \sum_{i=1}^{k} \alpha_i x_i$, where $\alpha_0,\ldots,\alpha_k \in \mathbb{R}$,
	and a finite set $A \subset \mathbb{R}$, determine whether there exists $(x_1, \ldots, x_k) \in A^k$ such that $\phi(x_1, \ldots, x_k)=0$.
	When $\phi$ is $\sum_{i=1}^k x_i$ the problem is called $k$-\SUM.
\end{description}
There are simple algorithms that solve $k$-\LDT{} in time $O(n^{(k+1)/2})$ when $k$ is odd, or $O(n^{k/2}\log n)$ when $k$ is even; see~\cite{AilonC05}.
These algorithms are based on straightforward reductions to a \TwoSUM{} problem or to an unbalanced \ThreeSUM{} problem,
depending on whether $k$ is even or odd, respectively.
These are currently the best known time upper bounds for solving $k$-\LDT{}.
Erickson~\cite{Erickson99} showed that, for an even $k$, there is a $k$-linear decision tree with depth $O(n^{k/2})$, removing an $O(\log n)$ factor when comparing to the uniform model.  
The above bounds match with the seminal lower bound results of Erickson~\cite{Erickson99}, and Ailon and Chazelle~\cite{AilonC05},
who showed that any $k$-linear decision tree for solving $k$-\LDT{} must have
depth $\Omega(n^{k/2})$ when $k$ is even and $\Omega(n^{(k+1)/2})$ when $k$ is odd.
In particular, any 3-linear decision tree for \ThreeSUM{} has depth $\Omega(n^2)$.
\GP{} showed that using more variables per comparison leads to a dramatic improvement in the depth of the tree, which significantly beats the above lower bounds.
Specifically, as will be reviewed below, they showed  that there is a 4-linear decision tree for \ThreeSUM{} with depth $O(n^{3/2}\sqrt{\log n})$, and by the reduction from $k$-\LDT{} to unbalanced \ThreeSUM{},
they concluded that there is a $(2k-2)$-linear decision tree for $k$-\LDT{} with depth $O(n^{k/2}\sqrt{\log n})$, for any odd $k\ge 3$.
If we allow arbitrarily many variables in a comparison (i.e., not a constant),
then for large $k$ (e.g., $k\geq 7)$, the decision tree complexity for $k$-\SUM{} goes down even more drastically;
see~\cite{CIO16, ES16}, for recent improvements on such settings.

Apart from the many lower bounds obtained from the conjectured hardness of \ThreeSUM{} and its variants, in recent years,
many lower bounds were obtained also from two other plausible conjectures.
The first is that computing the $(\min,+)$--product of two $n\times n$ matrices takes $\Omega(n^{3-o(1)})$ time (aka APSP-Hardness);
see for examples~\cite{WilliamsW10, AW15, AbboudW14}.
The second is that CNF-SAT takes $\Omega(2^{(1-o(1))n})$ time. The latter is often referred to as the {\em Strong Exponential Time Hypothesis} (SETH)~\cite{Impagliazzo2001, IPZ01}.
A natural question is whether any of these conjectures (\ThreeSUM{}, SETH, APSP) are in fact equivalent, or whether they all derive from a basic unifying hypothesis.
At the current state of knowledge, there is no strong relationship between any pair of these problems,
so it may be possible that the any one of them could be true or false, independently of the status of the others.
A recent breakthrough by Carmosino, Gao, and Impagliazzo~\cite{CarmosinoGIMPS15} provides evidence that such a relationship is {\em unlikely},
based on a nondeterministic variant of SETH; see~\cite{CarmosinoGIMPS15} for details.

\subsection{Our Results}
The following theorems capture our main results.

\begin{theorem}\label{thm:3SUM-DEC}
	The randomized $4$-linear decision tree complexity of \textnormal{\ThreeSUM{}} is $O(n^{3/2})$.
\end{theorem}

\begin{theorem}\label{thm:kLDT}
	The randomized $(2k-2)$-linear decision tree complexity of $k$-\textnormal{\SUM{}} and of $k$-\textnormal{\LDT{}} is $O(n^{k/2})$, for any odd $k\ge 3$. 
\end{theorem}
Theorems~\ref{thm:3SUM-DEC} and~\ref{thm:kLDT} improve (albeit randomized) the respective 
$O(n^{3/2}\sqrt{\log n})$-depth and $O(n^{k/2}\sqrt{\log n})$-depth decision trees given by \GP{}.
Our new decision tree bounds for $k$-\LDT{} (and $k$-\SUM{}) match the corresponding decision tree bound of Erickson~\cite{Erickson99} for {\em even} $k$.
We leave it as an open question whether there exist an $O(1)$-linear decision tree with depth $o(n^{3/2})$ for \ThreeSUM{} (see discussion in Section~\ref{sec:dis}).

Our technique includes some new insights on the $\ThreeSUM{}$ problem, and a specialized data structure, based
on an unusual randomized variant of fractional cascading in a grid.

As a supplement, we give (in the Appendix) an actual deterministic algorithm for \ThreeSUM{} that runs in $O(n^2 \log \log n / \log n)$ time.\footnote{
We consider a simplified Real RAM model.  Real numbers are subject to only two unit-time operations: addition and comparison.
In all other respects the machine behaves like a $w=O(\log n)$-bit word RAM with the standard repertoire of unit-time 
$AC^0$ operations: bitwise Boolean operations, left and right shifts, addition, and comparison.	}
The latter improves the $O(n^2 (\log\log n / \log n)^{2/3})$-time bound of \GP{}, 
and matches the bound given by a recent independent work of Freund~\cite{F15}.
Both algorithms, Freund's~\cite{F15} and ours, have common high-level ideas,
but ours makes a better use of word-RAM model, and is hence somewhat simpler.\footnote{
The independent result of Freund~\cite{F15} was brought to our attention after the completion of an initial version of this paper; see~\cite{GoldS15}.}

Recently, Lincoln, Vassilevska-Williams, Wang, and Williams~\cite{LWWW16} showed a reduction result in which
they apply our \ThreeSUM{} algorithm (based on an initial version of this paper~\cite{GoldS15}) as a black-box, leading to
a \ThreeSUM{} algorithm that uses only $O\left(\sqrt{n\log n / \log\log n} \right)$ space,
while preserving the time bound of our algorithm.

\section{Methods and Lemmas}\label{sec:lemmas}
We give an overview of the techniques discussed above.
This includes works by Fredman~\cite{Fredman76} and Chan~\cite{Chan08}.
In some of our results we will use a special randomized variant of {\em fractional cascading}
(Chazelle and Guibas~\cite{CG86-1, CG86-2}), which we also review here.
This will later allow us to review the results of \GP{} and to obtain our new results.

Throughout the paper we refer to the trivial (albeit ingenious) observation that $a+b < a'+b'$ iff $a-a' < b'-b$ as \emph{Fredman's trick}.
We denote by $[N]$ the first $\ceil{N}$ natural numbers succeeding zero $\{1,\ldots, \ceil{N} \}$, where $N$ may or may not be an integer.

Fredman showed that, given $n$ numbers whose sorted order is one of $\Pi \le n!$ realizable permutations,
they can be sorted using a linear number of comparisons when $\Pi$ is sufficiently small.
More generally, we have:
\begin{lemma}[Fredman 1976~\cite{Fredman76}]\label{lem:Fredman}
	A list $L$ of $n$ numbers, whose sorted order is one of $\Pi$ possible permutations, can be sorted with
	$2n+\log\Pi$ pairwise comparisons.
\end{lemma}

\paragraph{Sorting Pairwise Sums and its Geometric Interpretation.}
Fredman describes the relation between the complexity of hyperplane arrangements and
the decision tree complexity of sorting pairwise sums.
\GP{} use similar arguments in their \ThreeSUM{} decision tree where they sort pairwise sums.
Given lists $A = (a_i)_{i\in[n]}$ and $B=(b_i)_{i\in [n]}$ of distinct real numbers,
define the pairwise sum $A+B = \{a_i + b_j \:|\: i,j \in [n]\}$.
The input $A, B$ can be regarded as a point  $p = (a_1,\ldots,a_n,b_1,\ldots,b_n)\in \mathbb{R}^{2n}$.
The points in $\mathbb{R}^{2n}$ that agree with a fixed permutation of $A+B$ form a convex cone bounded
by the set $H$ of the $\binom{n^2}{2}$ hyperplanes $x_i + y_j - x_k - y_l = 0, \mbox{ for } i,j,k,l\in[n]$, $(i,j)\neq(k,l)$.
The number of possible sorted orders of $A+B$ is therefore bounded by the number of regions
(of all dimensions) in the arrangement $\A(H)$ of $H$.
As shown by Buck~\cite{Buck43}, the number of regions in an arrangement of $m$ hyperplanes in $\mathbb{R}^d$ of dimension $k\le d$ is at most
\[
\binom{m}{d-k}\left(\binom{m - d+k}{0} + \binom{m - d+k}{1} + \cdots + \binom{m - d+k}{k}\right).
\]
Thus, the number of regions of all dimensions is $O(m^d)$ (where the constant of proportionality is independent of $d$).
Hence, the number of possible sorting permutations of $A+B$ is $O\left((n^{4})^{2n}\right) = O(n^{8n})$.
One can also construct the hyperplane arrangement explicitly in $O(m^d)$ time
by a standard incremental algorithm~\cite{EdelsbrunnerOS86}.
The following lemma, taken from \GP{}, extends this analysis by considering only a subset of these hyperplanes,
and is an immediate consequence of these observations.

\begin{lemma}\label{lem:sortX+Y}
	Let $A=(a_i)_{i\in[n]}$ and $B=(b_i)_{i\in [n]}$ be two lists, each of $n$ real numbers,
	and let $F\subseteq [n]^2$ be a set of positions in the $n\times n$ grid.
	The number of realizable orders of
	$(A+B)_{|F} := \{a_i + b_j \:|\: (i,j) \in F\}$ is $O\mathopen{}\left({\binom{|F|}{2}}^{2n}\right)\mathclose{}$,
	and therefore $(A+B)_{|F}$ can be sorted with at most $2|F| + 4n\log|F| + O(1)$ comparisons.
\end{lemma}
In Lemma~\ref{lem:sortX+Y}, the case $F=[n]^2$ goes back to Fredman~\cite{Fredman76}, who showed that $O(n^2)$
comparisons suffice to sort $A+B$.

For some of the algorithms presented and reviewed in this paper, it is important to assume that the elements of the pairwise sum
are distinct, and therefore have a unique sorting permutation. When numbers do
appear multiple times, a unique sorting permutation can be obtained by breaking ties consistently (see~\cite{GP14} for details).

\paragraph{Iterative Search and Fractional Cascading.}\label{subsec:fractional}
In our decision tree construction for \ThreeSUM{},
we aim to speed-up binary searches of the same number, in many sorted sets.
We will use for this task a special randomized variant of {\em fractional cascading}, which will be described in Section~\ref{sec:Randomized}.
First, we briefly recall the standard fractional cascading technique, which was introduced
by Chazelle and Guibas~\cite{CG86-1, CG86-2}, for solving the {\em iterative search problem}, defined as follows.
Let $U$ be an ordered universe of keys.
Define a {\em catalog} as a finite ordered subset of $U$.
Given a set of $k$ catalogs $C_1, C_2,\ldots, C_k$ over $U$, such that $|C_i|=n_i$ for each $i\in [k]$,
and $\sum_{i=1}^{k} {n_i} = n$, the iterative search problem is to provide a data structure that supports efficient execution of queries of the form:
given a query $x\in U$, return the largest value less than or equal to $x$ in each of the $k$ catalogs.

Fractional cascading lets one preprocess the catalogs in $O(n)$ time, using $O(n)$ storage,
and answer iterative search queries in $O(\log n + k)$ time per query. This is essentially optimal in terms of query time, storage size and preprocessing time.
The idea is to maintain a sufficient number of pointers across catalogs, so that,
once we have the answer $c_i$ to a query in a catalog $C_i$, we can follow a pointer to an element in $C_{i+1}$,
which is only $O(1)$ indices away from the answer $c_{i+1}\in C_{i+1}$.

In order to obtain optimal query time, the fractional cascading method expands each catalog $C_i$ to an augmented catalog $L_i$,
starting with $L_k$ and proceeding backwards down to $L_1$.
$L_k$ is the same as $C_k$, and for each $1 \leq i < k$, $L_i$ is formed by merging $C_i$ with every second element of $L_{i+1}$. The items in $C_i$ that were not originally in the catalog are marked as synthetic keys.
From each synthetic key in $C_i$ we add a bridge (pointer) to the element in $L_{i+1}$ on which it was based.
Using these bridges and additional pointers, from each real key to the two consecutive synthetic keys nearest to it, one can follow directly from each element of $L_i$ (real or synthetic) to the elements in $L_{i+1}$ nearest to it,
and by construction, the gap between these elements is $2$. 
Thus, given a query number $x$, after spending $O(\log n)$ time for searching it in $L_1$, it takes only $O(1)$ time to locate $x$ in each subsequent catalog,
for a total of $O(\log n +k)$ time, as described.

Fractional cascading can also be extended to support 
a collection of catalogs stored at the vertices of a directed acyclic graph (DAG),
and each query searches with some specified element $x$ through the catalogs stored at the nodes of some specified path in the DAG.
In more detail, a {\em catalog graph} is a DAG in which each vertex stores a catalog (ordered list of keys).
A query consists of a key $x$ and a path $\pi$ in the graph, and the goal is to search with $x$
in the catalog of each node of $\pi$.
When the maximum in/out degree $\Delta$ of the catalog graph is constant,
fractional cascading can be extended to this scenario, with the same bounds as before (albeit with larger constants of proportionality).
Here too each catalog $C_v$ at a node $v$, is expanded into an augmented catalog $L_v$,
and each $L_v$ passes to its predecessors every $2\Delta$-th element (instead of every second element in the earlier case, where $\Delta$ was $1$).
See~\cite{CG86-1, CG86-2} for more details on the construction of the data structure, proof of correctness, and performance analysis.

In our algorithms we will present a special non-standard variant of this method,
that lets us preserve the advantages of the other techniques (most notably, Fredman's trick) that we use.

\paragraph{Bichromatic Dominance Reporting.}
Given a finite set $P$ of red points and blue points in $\mathbb{R}^d$,
the {\em bichromatic dominating pairs} problem is
to enumerate all the pairs $(p,q)\in P^2$ such that $p$ is red, $q$ is blue, and $p$ dominates $q$, i.e.,
$p$ is greater than or equal to $q$ at each of the $d$ coordinates.
A natural divide-and-conquer algorithm~\cite[p.~366]{PreparataShamos85} runs in $O(|P|\log^d |P| + K)$ time,
where $K$ is the output size.  Chan~\cite{Chan08} provided an improved {\em truly subquadratic} time bound (excluding the cost of reporting the output) when $d=O(\log |P|)$,
with a sufficiently small constant of proportionality.
\begin{lemma}[Chan~\cite{Chan08}]\label{lem:redblue}
	Given a finite set $P\subset \mathbb{R}^d$ of red and blue points, one can compute
	all bichromatic dominating pairs $(p,q)\in P^2$
	in time  $O(c_\epsilon^d |P|^{1+\epsilon} + K)$, where $K$ is the output size,
	$\epsilon \in (0,1)$ is arbitrary, and $c_\epsilon = 2^\epsilon/(2^{\epsilon}-1)$.
\end{lemma}
Throughout the paper, we invoke Lemma~\ref{lem:redblue} a large number of times, with $\epsilon=1/2, c_\epsilon \approx 3.42,$ and
$d = \delta \log n$, where $\delta >0$ is sufficiently small
to make the overall running time of all the invocations dominated by the total output size; see below (Appendix) for details.

\paragraph{The Quadratic \ThreeSUM{} Algorithm.}\label{subsec:quadratic}

We next give a brief overview of the quadratic-time algorithm. We follow the implementation given by \GP{},
which is slightly different from the standard approach, but is useful for the explanation of the subquadratic algorithms in~\cite{GP14} and in this paper.
For later references, we present the algorithm for the more general three-set version of \ThreeSUM{}. We recall that in this setup we have
three finite sets $A,B,C\subset \mathbb{R}$, and the problem is to determine whether
there exist $a\in A, b\in B, c\in C$ such that $a+b+c=0$.
If the sets $A,B,C\subset \mathbb{R}$ are not all of the same size, the problem is named unbalanced \ThreeSUM{}, as noted earlier.

The algorithm runs over each $c\in C$ and searching for $-c$ in the pairwise sum $A+B$.
With a careful implementation, given below,
each search takes $O(|A|+|B|)$ time, for a total of $O(|C|(|A|+|B|))$ time.  We view $A+B$ as being a matrix
whose rows correspond to the elements of $A$ and columns to the elements of $B$, both listed in increasing order.
To help visualizing some steps of the algorithms, we think of the rows arranged in increasing order from top
to bottom, and of the columns from left to right, see, e.g., Figure~\ref{fig:algorithm} (in the Appendix).

\setdescription{leftmargin=1.5cm,labelindent=\parindent}
\begin{description}\label{alg:quadratic}
	\addtolength{\itemsep}{-0.35\baselineskip}
	\item[1.$\;$] Sort $A$ and $B$ in increasing order as $A(0),\ldots, A(|A|-1)$ and $B(0),\ldots, B(|B|-1)$.
	\item[2.$\;$] For each $c\in C$,
	\item [2.1.$\;\;$] Initialize $\lo \leftarrow 0$ and $\hi \leftarrow |B|-1$.
	\item [2.2.$\;\;$] Repeat:
	\item [2.2.1.$\;\;\;\;$] If $-c = A(\lo) + B(\hi)$, report witness ``$(A(\lo), B(\hi), c)$''.
	\item [2.2.2.$\;\;\;\;$] If $-c > A(\lo) + B(\hi)$ then increment $\lo$, otherwise decrement $\hi$.
	\item [2.3.$\;\;\;$] Until $\lo = |A|$ or $\hi = -1$.
	\item[3.$\;$] If no witnesses were found report ``no witness.''
\end{description}

The correctness easily follows from the fact that each row and column of $A+B$ is sorted in increasing order.
Note that when a witness is discovered in Step 2.2.1, the algorithm can stop right there.
However, in order to simplify future definitions and explanations, this implementation continues to search for more witnesses.
For our purpose, after finding a witness we will always choose to decrement $\hi$. This choice will be made throughout the paper.

Define the {\em contour} of $x$, $\contour(x,A+B)$, (\contour(x), when the context is clear) to be the sequence of positions $(\lo,\hi)$
encountered while searching for $x$ in $A+B$ in the preceding algorithm.
Lemma~\ref{lem:contour} is straightforward.

\begin{lemma}\label{lem:contour}
	For $x < y\in \reals$, \contour(x) lies fully above \contour(y);
	that is, for each $i, i', j\in \{0,\ldots, n-1 \}$, if $(i, j)\in \contour(x)$ and $(i', j )\in \contour(y)$,
	then $i\leq i'$.
\end{lemma}
By Lemma~\ref{lem:contour} a pair of contours can overlap, but never cross.
Moreover, Lemma~\ref{lem:contour} implies a weak total order relation $\prec$ on the contours, which corresponds to the order between the searched elements,
such that  $x < y$ iff $\contour(x) \prec \contour(y)$, where the latter relation means that the two contours satisfy the properties stated in the lemma;
see Figure~\ref{fig:algorithm} (in the Appendix).

\section{Gr\o nlund and Pettie's Subquadratic \ThreeSUM{} Decision Tree}\label{sec:breakthrough}
In this section we give an overview of the subquadratic decision tree of Gr\o nlund and Pettie~\cite{GP14}.
In the following sections we show how their ideas can be extended and combined with additional techniques,
to yield our improved results.

\subsection{Subquadratic Decision Tree}\label{subsec:decision-tree}
We give an overview of the subquadratic decision tree for \ThreeSUM{} over a single input set $A$ of size $n$, taken from~\cite{GP14},
resulting in a $4$-linear decision tree with depth $O(n^{3/2}\sqrt{\log n})$.
This is shown by an algorithm that performs at most $O(n^{3/2}\sqrt{\log n})$ comparisons, where each comparison is a sign test of a linear expression with at most 4 terms.

\setdescription{leftmargin=1.5cm,labelindent=\parindent}
\begin{description}
	\addtolength{\itemsep}{-0.2\baselineskip}
	\item[1.$\;$] Sort $A$ in increasing order as $A(0),\ldots, A(n-1)$.
	Partition $A$ into $\ceil{n/g}$ groups $A_1,\ldots, A_{\ceil{n/g}}$, each of at most $g$ consecutive elements,
	where $g$ is a parameter that we will fix later, by setting
	$A_i := \{A((i-1)g), \ldots, A(ig-1)\}$, for each $i=1,\ldots, \ceil{n/g}-1$, where $A_{\ceil{n/g}}$ may be smaller.
	The first and last elements of $A_i$ are $\min(A_i) = A((i-1)g)$ and $\max(A_i) = A(ig-1)$.
	
	\item[2.$\;$] Sort $D := \bigcup_{i\in [n/g]} \left(A_i-A_i\right) = \{a - a' \;|\; a,a' \in A_i \mbox{ for some } i\}$.
	\item[3.$\;$] For all $i,j \in [n/g]$, sort $A_{i,j} := A_i + A_j = \{a + b \;|\; a\in A_i \mbox{ and } b\in A_j\}$.
	\item[4.$\;$] For $k$ from 1 to $n$,
	\item [4.1.$\;\;$] Initialize $\lo \leftarrow 1$ and $\hi \leftarrow \ceil{n/g}$.
	\item [4.2.$\;\;$] Repeat:
	\item [4.2.1.$\;\;\;\;$] If $-A(k) \in A_{\lo,\hi}$, report ``solution found'' and halt.
	\item [4.2.2.$\;\;\;\;$] If $\max(A_{\lo}) + \min(A_{\hi}) > -A(k)$ then decrement $\hi$, otherwise increment $\lo$.
	\item [4.3.$\;\;\;$] Until $\lo = \ceil{n/g}+1$ or $\hi = 0$.
	\item[5.$\;$] Report ``no solution'' and halt.
\end{description}
This algorithm can be generalized in a straightforward way to solve the (unbalanced) three-set version of \ThreeSUM{}.
For the easy argument concerning the correctness of the algorithm, see~\cite{GP14}.

With a proper choice of $g$, the decision tree complexity of the algorithm is $O(n^{3/2}\sqrt{\log n})$.
Step 1 requires $O(n\log n)$ comparisons.
By Lemma~\ref{lem:sortX+Y},
Step 2 requires $O(n\log n + |D|) = O(n\log n + gn)$ comparisons to sort $D$.
By Fredman's trick, if $a,a' \in A_i$ and $b,b'\in A_j$, $a + b < a' + b'$ holds iff $a-a' < b'-b$,
and both sides of this inequality are elements of $D$. Thus, Step 3 does not requires any real input comparisons, given the sorted order on $D$.
For each iteration of the outer loop (in Step 4) there are at most $2\ceil{n/g}$ iterations of the inner loop (Step 4.2), since each iteration
ends by either incrementing $\lo$ or decrementing $\hi$.
In Step 4.2.1 we can determine whether $-A(k)$ is in $A_{\lo,\hi}$
using binary search, in $\log|A_{\lo,\hi}| = O(\log g)$ comparisons.
The total number of comparisons is thus
$O(n\log n + gn + (n^2\log g)/g)$, which becomes $O(n^{3/2}\sqrt{\log n})$ when $g = \sqrt{n\log n}$.

\section{Improved Decision Trees for \ThreeSUM{}, $k$-\SUM{}, and $k$-\LDT{}}\label{sec:Randomized}
In this section we show that the randomized decision tree complexity of \ThreeSUM{} is $O(n^{3/2})$,
and more generally, that the randomized decision tree complexity of $k$-\LDT{} is $O(n^{k/2})$, for any odd $k\geq 3$.
This bound removes the $O(\sqrt{\log n})$ factor in Gr\o nlund and Pettie's decision tree bound.
We show this results by giving a randomized algorithm that constructs a $(2k-2)$-linear decision tree whose expected depth is $O(n^{k/2})$.

To make the presentation more concise, we present it for the variant where
we have three different sets $A$, $B$, $C$ of $n$ real numbers each, and we want
to determine whether there exist $a\in A$, $b\in B$, $c\in C$, such that $a+b+c=0$.

As in the previous section, we partition each of the sorted sets
$A$ and $B$ into $\ceil{n/g}$ blocks, each consisting of $g$ consecutive elements,
denoted by $A_1,\ldots,A_{n/g}$, and $B_1,\ldots,B_{n/g}$, respectively.
As above, but with a slightly different notation, we consider the $n\times n$
matrix $M=M^{AB}$, whose rows (resp., columns)
are indexed by the (sorted) elements of $A$ (resp., of $B$), so that
$M(k,\ell) = a_k+b_\ell$, for $k,\ell\in [n]$.
The partitions of $A$ and of $B$ induce, as before, a partition of $M$ into $n^2/g^2$
boxes $M_{i,j}$, for $i,j\in[n/g]$, where $M_{i,j}$ is the portion of $M$ with
rows in $A_i$ and columns in $B_j$.

Fredman's trick allows us to sort all the boxes $M_{i,j}$ with $O(ng)$ comparisons.
Since the problem is fully symmetric in $A$, $B$, $C$, we can also define
analogous matrices $M^{AC}$ and $M^{BC}$, constructed in the same manner for the
pairs $A$, $C$ and $B$, $C$, respectively, partition each of them into $n^2/g^2$ boxes,
and obtain the sorted orders of all the corresponding boxes, with $O(ng)$ comparisons.

The crucial (costliest) step in Gr\o nlund and Pettie's algorithm, which we are going
to improve, is the searches of the elements of $-C$ in $M^{AB}$.
For each $c\in C$, let $\sigma(c) = \contour(-c)$ denote the staircase path contour of $-c$, as defined before Lemma~\ref{lem:contour}.
The length of $\sigma(c)$ is thus at most $2n$.
Each of the paths $\sigma(c)$ visits some (at most $2\ceil{n/g}$) of the boxes $M_{i,j}$,
and the index pairs $(i,j)$ of these boxes also form a staircase pattern, as in the preceding sections.
The number of boxes that a contour $\sigma(c)$ visits is at most $2\ceil{n/g}$.
For each $c\in C$, the sequence of boxes that $\sigma(c)$ visits can be obtained by invoking (an appropriate variant of)
Step 4 of the algorithm in Section~\ref{subsec:decision-tree}, excluding the binary search in Step 4.2.1. The total running time of this step, over all $c\in C$, is $O(n^2 /g)$.

The paths $\sigma(c)$, being contours, have the structure given in Lemma~\ref{lem:contour}, including the weak total order $\prec$ between them.
As a corollary, we obtain:

\medskip

\noindent{\bf Claim:}
For each box $M_{i,j}$, let $C_{i,j}$ denote the set of elements of $C$ whose
paths $\sigma(c)$ traverse $M_{i,j}$. Then $C_{i,j}$ is a contiguous subsequence of $C$.

\medskip

Put $\kappa_{i,j} := |C_{i,j}|$. Then we clearly have
$
\sum_{i,j\in[n/g]} \kappa_{i,j} = O(n^2/g) .
$
That is, the average number of elements of $c$ that visit a box is $O(g)$, and, for
each box, these elements form a contiguous subsequence of $C$, as just asserted. Let $C^*_{i,j}$ denote
the contiguous sequence of indices in $C$ of the elements of $C_{i,j}$. That is,
$C_{i,j} = \{c_\ell \mid \ell\in C^*_{i,j} \}$.
With all these observations, we next proceed to derive the mechanism by which, for each box $M_{i,j}$,
we can efficiently search in $M_{i,j}$ with the (negations of the) $\kappa_{i,j}$ corresponding elements of $C_{i,j}$.

We apply a special variant of fractional cascading. The twist is in the way in which we construct the augmented catalogs.
Note that in each box $M_{i,j}$, we have $g^2$ elements of the form $a_{k} + b_{\ell}$, but only $2g$ indices $k, \ell$.
We want to sample elements from a box, and then copy and merge them into its neighbor boxes. However, in order to be able to use Fredman's trick,
we have to preserve the property that the number of element-indices (rows and columns) in each box
stays $O(g)$ (unlike a naive implementation of fractional cascading, where it is enough that each box be of size $O(g^2)$).

Thus, we sample elements from $A$ (row elements) and elements from $B$ (column elements) separately.
We construct augmented sets $A'_1, \ldots, A'_{\ceil{n/g}}$.
Starting with $A'_{\ceil{n/g}} = A_{\ceil{n/g}}$, we sample each element in $A'_{\ceil{n/g}}$ with probability $p=\frac{1}{4}$.
Each sampled element is copied and merged with $A_{\ceil{n/g}-1}$,
and we denote by $A'_{\ceil{n/g}-1}$ the new augmented set.
Then we sample each element from $A'_{\ceil{n/g}-1}$ with the same probability $p$, copy and merge the sampled elements with $A_{\ceil{n/g}-2}$, obtaining $A'_{\ceil{n/g}-2}$,
and continue this process until the augmented set $A'_1$ is constructed.
Similarly, we construct the augmented sets $B'_1, \ldots, B'_{\ceil{n/g}}$, but we do it in the opposite direction, starting from $B'_1 = B_1$ and ending with $B'_{\ceil{n/g}}$.
Clearly, similar to standard fractional cascading,
the expected size of each of the augmented sets is $O(g)$, as the expected numbers of additional elements placed in each box form a convergent geometric series.
Now we sort
\[
D_{A'} = \bigcup_{i\in [n/g]} \left(A'_i-A'_i\right) = \{a - a' \;|\; a,a' \in A'_i \mbox{ for some } i\}.
\]
In each $A'_i-A'_i$, the expected number of elements $a_{k} - a_{k'}$ is $O(g^2)$, and the expected number of element indices $k, k'$ is only $O(g)$.
Thus, by Lemma~\ref{lem:sortX+Y}, we can sort $D_{A'}$ with  expected $O(ng)$ comparisons.
Similarly, we sort $D_{B'} = \bigcup_{j\in [n/g]} \left(B'_j-B'_j\right)$ with the same expected number of comparisons.
Then, we form the union $D'=D_{A'} \cup D_{B'}$ and obtain its sorted order by merging $D_{A'}$ and $D_{B'}$. This costs additional expected $O(ng)$ comparisons.
By Fredman's trick, from the sorted order of $D'$, we can obtain the sorted order of the augmented boxes $A'_i + B'_j$, for each $i,j\in [n/g]$, without further comparisons.

With these augmentations of the row and column blocks,
the matrix $M^{AB}$ itself is now augmented, such that each modified box $M_{i,j} = A'_i + B'_j$
receives some fraction of the rows from the box $M_{i+1,j}$ below it, and a fraction of the columns from the box $M_{i, j-1}$ to its left.
Each box $M_{i,j}$ corresponds to a vertex in the catalog graph, and it has (at most) two outgoing edges, one to the vertex that corresponds to $M_{i+1,j}$
and one to the vertex that corresponds to $M_{i, j-1}$ (it also has at most two incoming edges).
Clearly this is a DAG with maximum in/out degree $\Delta = 2$, which is why we sampled $\frac{1}{2\Delta} = \frac{1}{4}$ of the rows/columns in each step.
We complete the construction of this special fractional cascading data structure, by adding the appropriate pointers, similar to what is done in a standard implementation of fractional cascading (see Section~\ref{subsec:fractional}).
This does not require any further comparisons, since the pointers from synthetic keys (the sampled elements) to real keys, and pointers from real keys to synthetic keys,
depend only on the sorted order of the augmented sets $M_{i,j}$, which we already computed.
So the overall expected number of comparisons needed to construct this data structure is still $O(ng)$.

Consider now the search with $-c$, for some $c\in C$.
Assume that the search has just visited some box $M_{i',j'}$,
and now proceeds to search in box $M_{i,j}$. Thus, either
$(i,j)=(i'+1,j')$ or $(i,j)=(i',j'-1)$. Assume, without loss of generality, that
$(i,j)=(i'+1,j')$; a symmetric argument applies when $(i,j)=(i',j'-1)$, using columns instead of rows.
In this case, the fractional cascading mechanism has sampled, in a random manner,
an expected quarter of the rows of (the already augmented) $M_{i,j}$ and has sent them to $M_{i',j'}=M_{i-1,j}$.
The output of the search at $M_{i-1,j}$, if $-c$ was not
found there, includes two pointers to the largest element $\xi^-$ of $M_{i,j}$
that is smaller than $-c$, and to the smallest element $\xi^+$ of $M_{i,j}$ that
is larger than or equal to $-c$.
We need to go over the elements in the sorted order of $M_{i,j}$ that lie between
$\xi^-$ and $\xi^+$, and locate $-c$ among them. If we do not find it, we get the
two consecutive elements that enclose $-c$, retrieve from them two corresponding
pointers to a pair of elements in the next box to be searched,
that enclose $-c$ between them, and continue the fractional cascading search in
the next box, in between these elements.

The main difficulty in this approach is that the number of elements of $M_{i,j}$
between $\xi^-$ and $\xi^+$ might be large,
because there might be many elements between $\xi^-$ and $\xi^+$ in rows that we did not sample, and then we have to inspect them all, slowing down the search.

Concretely, in this case we sample, in expectation,
a quarter of the rows of $M_{i,j}$ (recall that, we actually sample the rows from an augmented
box that has already received data from previous boxes, but let us ignore this issue for now).
Collectively, these rows contain (in expectation) $\Theta(g^2)$
elements of $M_{i,j}$, but we have no good control over the size of the gaps of
non-sampled elements between consecutive pairs of sampled ones.
This is because there might be rows that we did not sample which contain many elements between $\xi^-$ and $\xi^+$, 
and searching through such large gaps could slow down the procedure considerably.
See Figure~\ref{fig:dectree} for an illustration.
(For a normal fractional cascading, this would not be an issue, but here the peculiar and implicit
way in which we sample elements has the potential for creating this problem.)

\newcommand{\cellgrey}{\cellcolor{gray!35!}}
\begin{figure}[tb]
	\centering
	\begin{tabular}{|c|c|c|c|c|c|c|c|c|c|}
		\hline
		\cellgrey 60 & \cellgrey 70 & \cellgrey 80 & \cellgrey 90 & \cellgrey 100 & \cellgrey 110 & \cellgrey 120 & \cellgrey 130 & \cellgrey 140 & \cellgrey 150  \\
		\hline
		160 &  170 &  180 &  190 &  200 & 210 & 220 & 230 & 240 & 250 \\
		\hline
		\cellgrey 260 & \cellgrey 270 & \cellgrey 280 & \cellgrey 290 & \cellgrey 300 & \cellgrey 310 & \cellgrey 320 & \cellgrey 330 & \cellgrey 340 & \cellgrey 350 \\
		\hline
	\end{tabular}
	\caption{\label{fig:dectree}
		An expensive step in the fractional cascading search: Assume that only the first and third rows (appearing in gray) are sent to the preceding box (above the current one), 
		and that we search with $-c=205$. The previous search locates $-c$ between $\xi^- = 150$ and $\xi^+ = 260$, say, and now we have to examine the entire second row to locate $-c$ in the current box. 
	}
\end{figure}

We handle this problem as follows. Consider any gap of non-sampled elements of
$M_{i,j}$ between a consecutive pair $\xi^- < \xi^+$ of sampled ones. We claim
that the expected number of rows to which these elements belong is $O(1)$.
(Note that this is why we needed randomization; if we sampled every $4$th element in $A_i$ and $B_j$ deterministically then the rows-gap between $\xi^-$ to $\xi^+$ could be much larger, in all boxes $M_{i,j}$.) 
Indeed, the probability to have $k$ distinct rows in such a gap, conditioned on the choice of the row containing $\xi^-$, is $\frac{1}{4}\left(\frac{3}{4}\right)^k$,
which follows since each row is sampled \emph{independently} with probability $1/4$.
Hence, the (conditionally) expected row-size of a gap is 
$$\sum_{k\ge 0} k \frac{1}{4}\left(\frac{3}{4}\right)^k = O(1),$$
as claimed.
Denote this expected value as $\beta$.
In other words, for each $c\in C_{i,j}$, let $R_c$ be the set of rows that show up
in the gap between the corresponding elements $\xi^-$ and $\xi^+$ for $c$.
The overall expected size $\sum_{c\in C_{i,j}} |R_c|$ is thus $\beta|C_{i,j}|$.

Fix a box $M_{i,j}$. For each $\ell\in C^*_{i,j}$ and for each $k\in R_{c_\ell}$, we need
to locate $-c_\ell$ among the elements in row $k$ of $M_{i,j}$. That is, we need
to locate $-c_\ell$ among the elements of the set $a_k+B'_j$. This however is equivalent
to locating $-a_k-c_\ell$ among the elements of $B'_j$.

We therefore collect the set $S$ of all the sums $-a_k-c_\ell$, for $\ell\in C^*_{i,j}$
and $k\in R_{c_\ell}$, and recall that in expectation we have $|S| = O(|C_{i,j}|)$.
The crucial observation is that we already (almost) know the order of these sums.
To make this statement more precise, partition, in the usual manner, the sorted sequence $C$ into $\ceil{n/g}$
blocks $C_1,C_2,\ldots,C_{\ceil{n/g}}$, each consisting of $g$ consecutive elements in
the sorted order. As mentioned earlier, a symmetric application of Fredman's trick
allows us to obtain the sorted order of each box of the form $A'_i+C_j$, using a total
of $O(ng)$ comparisons.

The number of (consecutive) blocks $C_s$ of $C$
that overlap $C_{i,j}$ is $t_{i,j} \leq \lceil \kappa_{i,j}/g \rceil +2$. 
Moreover, each sum in $S$
belongs to $-(A'_i+C_s)$ for one of these $t_{i,j}$ blocks. Since each of these sets is
already sorted, we extract from them (with no extra comparisons) the elements of $S$ as the union of $t_{i,j}$ sorted
sequences $S_{i,s}$, where $S_{i,s} \subset -(A'_i+C_s)$ for each $s$. Arguing as above, the expected size of $S_{i,s}$ is $\beta|C_s| = O(g)$.
We now merge each of the sorted sequences $S_{i,s}$ with $B'_j$,
using an expected $O(g)$ comparisons for each merge.
As a result, each sum $-a_i-c_\ell$ is located between two consecutive elements
$b_{i,\ell}^- < b_{i,\ell}^+$ of $B'_j$.
In other words, for each $c_\ell\in C_{i,j}$, we have at most $|R_{c_\ell}|$ candidates
for being the largest element of $M_{i,j}$ that is smaller than $-c_\ell$
(these are the elements $a_i+b_{i,\ell}^-$, for $i\in R_{c_\ell}$), and we select the
largest of them, requiring no comparisons, as these are all elements of the
already sorted $A'_i+B'_j$. In the same manner, we find the smallest element of
$M_{i,j}$ that is larger than $-c_\ell$. Having found these two elements, we can
proceed to search $-c_\ell$ in the next box, using the appropriate pointers
created by the fractional cascading mechanism (see Section~\ref{sec:lemmas}).

The overall number of merges is
\[
\sum_{i,j\in[n/g]} t_{i,j} \le
\sum_{i,j\in[n/g]} \left( \kappa_{i,j}/g + 2 \right) = O(n^2/g^2),
\]
and each of them costs $O(g)$ expected comparisons, for a total of $O(n^2/g)$ expected comparisons.
Thus, the overall number of expected comparisons is $O(ng + n(\log g + n/g))$, 
which is $O(n^{3/2})$, when $g=\sqrt{n}$.
This completes the proof of Theorem~\ref{thm:3SUM-DEC}. \qed


\subsection{$k$-\SUM{} and Linear Degeneracy Testing}
The standard algorithm for $k$-variate linear degeneracy testing ($k$-\LDT) for odd $k\ge 3$, is based on a straightforward reduction to an instance of unbalanced \ThreeSUM{},
where $|A|=|B|=n^{(k-1)/2}$ and $|C|=n$; see~\cite{AilonC05} and~\cite{GP14}.
The analysis of this section also applies for unbalanced $\ThreeSUM{}$, and directly implies that it can be solved by using an expected number of
\[
O\left(g\left(|A| + |B| + |C|\right) +  |C|\left((|A|+|B|)/g + \log g\right)\right)
\]
comparisons, where the first term is the cost of sorting the blocks of (the augmented) $M^{AB}$, $M^{AC}$, and $M^{BC}$,
and where the second term is the cost of the fractional cascading searches.
We have
$|A|=|B|=n^{(k-1)/2}$, $|C|=n$, so by choosing $g=\sqrt{n}$, the bound becomes $O(n^{k/2})$.
Thus, the randomized decision tree complexity of $k$-\LDT{} (and thus of $k$-\SUM{}) is $O(n^{k/2})$, for any odd $k\ge 3$, as stated in Theorem~\ref{thm:kLDT}.

\section{Discussion}\label{sec:dis}
The exponent $3/2$ has a special significance in \Patrascu's framework~\cite{Patrascu10} of conditional lower
bounds based on hardness of \ThreeSUM{}, which was recently refined by Kopelowitz, Pettie, and Porat~\cite{KopelowitzPP14}.
Their lower bounds on triangle enumeration and polynomial lower bounds on dynamic data structures depend on the complexity of \ThreeSUM{} being
$O\left(n^{3/2 + \Omega(1)}\right)$. 
We conjecture that our new decision tree bounds are optimal in the $r$-linear decision tree model, where $r$ is a constant.

\bibliographystyle{abbrv}
\bibliography{3SUM}

\appendix

\section{Subquadratic \ThreeSUM{} Algorithms}
Before describing our subquadratic \ThreeSUM{} algorithm, we give a brief overview 
on the algorithm of \GP{}. We then extend their technique with some additional observations, 
yielding an improved deterministic subquadratic time bound.

\subsection{Overview of Gr\o nlund and Pettie's Subquadratic Scheme}\label{subsec:GPalgorithm}
\GP{} present two subquadratic algorithms for \ThreeSUM{},
one is relatively simple, and the second one has slightly faster runtime but is more involved.
Both algorithms are based on their decision tree algorithm outlined above,
except that they use a much smaller value of $g$,
in order to make the overall running time subquadratic.
We give here a brief overview of the simpler algorithm. Their second algorithm
has some common high-level features with our algorithm, presented
below, but our algorithm processes the data in a different, simpler, and more efficient manner.

Note that, sorting the set $D$ in the subquadratic decision tree lets one obtain a comparison-efficient way
to sort each of $A_{i,j}$.
However, the actual running time is even more than quadratic, when all operations are considered.
When the boxes $A_{i,j}$ are small enough, Gr\o nlund and Pettie showed that it is
possible to obtain the sorted orders in each of the $(n/g)^2$ boxes, in (all inclusive) subquadratic time.

Specifically, the algorithm enumerates {\em every}
permutation $\pi : [g^2]\rightarrow [g]^2$, where $\pi = (\pi_{\row},\pi_{\col})$ is decomposed into
row and column functions $\pi_{\row}, \pi_{\col} : [g^2] \rightarrow [g]$, so that $\pi(k) =(\pi_{\row}(k), \pi_{\col}(k))$, for each $k\in[g^2]$.
By definition, $\pi$ is the correct sorting permutation for the box $A_{i,j}$ iff
$A_{i,j}(\pi(t)) < A_{i,j}(\pi(t+1))$ for all $t\in [g^2-1]$.
Since $A_{i,j} = A_i+A_j$ this inequality can
also be written $A_i(\pi_{\row}(t)) + A_j(\pi_{\col}(t)) < A_i(\pi_{\row}(t+1)) + A_j(\pi_{\col}(t+1))$.
By Fredman's trick this is equivalent to saying that the (red) point $p_j$ dominates the (blue) point $q_i$, where
\begin{align*}
p_{j} &= \left(A_j(\pi_{\col}(2)) - A_j(\pi_{\col}(1)), \,\ldots\,, A_j(\pi_{\col}(g^2)) - A_j(\pi_{\col}(g^2-1))\right)\\
q_{i} &= \left(A_i(\pi_{\row}(1)) - A_i(\pi_{\row}(2)), \,\ldots\,, A_i(\pi_{\row}(g^2-1)) - A_i(\pi_{\row}(g^2))\right).
\end{align*}

Invoking $(g^2)!$ times the bichromatic dominance reporting algorithm from Lemma~\ref{lem:redblue}, we find,
for each $\pi$, all such dominating pairs, that is, all boxes $A_{i,j}$ sorted by $\pi$.
Note that, for each pair of indices $j, i$, there is exactly one
invocation of the dominating pairs procedure in which the corresponding points $p_j$ and $q_i$ are such that
$p_j$ dominates $q_i$; this follows because we assume that all elements of $A_{i,j}$ are distinct
(see a previous remark concerning this issue).
This is important in order to keep the overall output size subquadratic.

By Lemma~\ref{lem:redblue} and the remarks just made,
the time to report all red/blue dominating pairs, over all $(g^2)!$ invocations of the procedure,
is $O\mathopen{}\left((g^2)! c_\epsilon^{g^2-1} (2n/g)^{1+\epsilon} + (n/g)^2\right)\mathclose{}$, where the last term is the total size of the outputs (one for each box $A_{i,j}$).
For $\epsilon=1/2$ and $g = \frac{1}{2}\sqrt{\log n/\log\log n}$,
the first term turns out to be negligible.  The total running time is therefore
$O((n/g)^2)$ for dominance reporting,
and $O(n^2\log g/g) = O\mathopen{}\left(n^2 (\log\log n)^{3/2}/(\log n)^{1/2}\right)\mathclose{}$
for the binary searches in Steps 4.1--4.3.
By Lemma~\ref{lem:sortX+Y} and Fredman~\cite{Fredman76},
there are at most $O(g^{8g})$ realizable permutations of $A_{i,j}$ (which is much smaller than $(g^2)!$). Hence,
this algorithm can be slightly improved to run in
$O\mathopen{}\left(n^2 \log\log n / \sqrt{\log n} \right)\mathclose{}$ time,
by constructing the arrangement of the hyperplanes (as defined in Section~\ref{sec:lemmas}) explicitly, extracting from it the relevant permutations,
and choosing $g=\Theta(\sqrt{\log n})$.

\subsection{Improved Deterministic Subquadratic \ThreeSUM{} Algorithm}\label{sec:algorithm}

In the algorithm of Gr\o nlund and Pettie, described above, the boxes $A_{i,j}$ are sorted by
using Fredman's trick to transform each permutation into a sequence of $g^2-1$ comparisons,
which are then resolved by the bichromatic dominance reporting algorithm.
Consequently, the space into which these sequences are encoded is of dimension $g^2-1$,
thus having the $c^{g^2-1}_{\epsilon}$ factor in the running time of the bichromatic dominance reporting algorithm
forced us to use $g=\Theta(\sqrt{\log n})$. In order to use a larger value of $g$, we
want to reduce the dimension of the points. Thus, we want to find a method to sort smaller sets,
while still be able to do the binary searches in each box in $O(\log g)$ time.

Fix some $k\in [g^2]$, and let $(l,m)\in [g]^2$ be a point in the $g\times g$ grid, such that $A_{i,j}(l,m)$ is the $k$-th smallest element in the box $A_{i,j}$.
Let $\tau=(\tau_{\row},\tau_{\col})$ denote $\contour(A_{i,j}(l,m))$, and enumerate its elements as $\tau(1), \tau(2),\ldots$.
Recall that,
if $\tau(t+1) = \tau(t) + (0,-1)$ then $A_{i,j}(l,m) \leq A_{i,j}(\tau(t))$, otherwise;
if $\tau(t+1) = \tau(t) + (1,0)$ then $A_{i,j}(l,m) > A_{i,j}(\tau(t))$.
The contour starting position is $(\tau_{\row}(0),\tau_{\col}(0))= (1,g)$, 
and it ends at the first $t^*$ for which $\tau(t^*) = (g+1, \cdot)$ or $\tau(t^*) = (\cdot, 0)$.
Recall that a pair of contours \contour(x) and \contour(y) in $A_{i,j}$ may overlap, but can never cross; see Lemma~\ref{lem:contour}.

Let $\tau'(A_{i,j}(l,m)) = (\tau'(0), \tau'(1), \ldots, \tau'(t_{\tau'})) \subseteq \tau = \contour(A_{i,j}(l,m))$
be the {\em partial contour} of $\tau$, defined as the subsequence of positions of
$\tau$ at which we chose to go down (i.e., increment $\lo$ in
the quadratic algorithm). The sequence $\tau'(A_{i,j}(l,m))$ is of length at most $g$,
since it contains at most one element of each row (at which we go down, by incrementing $\lo$);
see Figure~\ref{fig:algorithm}.

\newcommand{\cellred}{\cellcolor[rgb]{1,0,0}}
\newcommand{\cellsky}{\cellcolor{-red!80!}}
\newcommand{\cellskylight}{\cellcolor{-red}}
\newcommand{\cellblue}{\cellcolor[rgb]{0,.5,1}}
\newcommand{\cellgreen}{\cellcolor[rgb]{0,1,0}}
\newcommand{\cellyellow}{\cellcolor[rgb]{1,1,0}}
\newcommand{\cellpurple}{\cellcolor{blue!35!}}
\newcommand{\cellpurplelight}{\cellcolor{blue!18!}}

\begin{figure}[tb]
	\centering
	\begin{tabular}{|c|c|c|c|c|c|c|c|c|c|}
		\hline
		372 &  389 &  407 &  439 &  454 &  480 & \cellpurplelight 534 & \cellpurple 609 & \cellpurple 635 & \cellgreen 655 \\
		\hline
		397 &  414 &  432 &  464 &  479 & \cellpurplelight 505 & \cellpurple 559 &  634 &  660 & \cellskylight 680 \\
		\hline
		420 &  437 &  455 &  487 &  502 & \cellpurplelight 528 &  582 &  657 &  683 & \cellskylight 703 \\
		\hline
		\  442 &  459 &  477 &  509 &  524 & \cellpurplelight 550 &  604 &  679 & \cellskylight 705 & \cellsky 725 \\
		\hline
		478 &  495 &  513 & \cellpurplelight 545 & \cellpurple 560 & \cellpurple 586 & \cellskylight 640 & \cellsky 715 & \cellsky 741 &  761 \\
		\hline
		500 &  517 & \cellpurplelight 535 & \cellpurple 567 &  582 &  608 & \cellskylight 662 &  737 &  763 &  783 \\
		\hline
		523 & \cellpurplelight 540 & \cellpurple \textbf{558} &  590 &  605 &  631 & \cellskylight 685 &  760 &  786 &  806 \\
		\hline
		\cellpurplelight 548 & \cellpurple 565 &  583 &  615 &  630 & \cellskylight 656 & \cellsky \textbf{710} &  785 &  811 &  831 \\
		\hline
		\cellpurple 594 &  611 &  629 &  661 &  676 & \cellskylight 702 &  756 &  831 &  857 &  877 \\
		\hline
		627 &  644 &  662 &  694 & \cellskylight 709 & \cellsky 735 &  789 &  864 & 890  & 910 \\
		\hline
	\end{tabular}
	\caption{\label{fig:algorithm}
		The sky-blue colored entries form \contour(710), and the purple colored ones form \contour(558);
		A shared cell is shown in green.
		The lighter colors (light purple and light sky-blue) depict their {\em partial contour}, that is, the positions of the contours where we chose to go down.
		All the elements in the matrix whose values are in $[558,710)$ are enclosed between these two contours,
		excluding the partial contour of 558 and including the partial contour of 710.}
\end{figure}

Since the rows of $A_{i,j}$ are sorted, each position $(a, b) \in \tau'(A_{i,j}(l,m))$ satisfies
\begin{align}
A_{i,j}(a ,b') < A_{i,j}(l, m) \quad \text{ for every } &b' \leq b.\label{eq:large} \\
A_{i,j}(a ,b'') \geq A_{i,j}(l ,m) \quad \text{ for every } &b'' > b.\label{eq:small}
\end{align}
Thus, $\tau'(A_{i,j}(l,m))$ partitions $A_{i,j}$ into two sets:
$A_{(i,j)L}\subset (-\infty,\, A_{i,j}(l,m))$
consists of the elements at positions in $\{(a,b') \;|\; (a,b)\in \tau'(A_{i,j}(l,m))\;  \text{and} \; b' \leq b\}$,
and $A_{(i,j)R} \subset [A_{i,j}(l,m), \infty)$
consists of the elements at positions in $\{(a,b'') \;|\; (a,b)\in \tau'(A_{i,j}(l,m))\; \text{and} \; b'' > b\}$.
Rows succeeding the last row of $\tau'(A_{i,j}(l,m))$ are fully contained in $A_{(i,j)R}$.
By construction, $A_{(i,j)L}$ is the set of all elements in $A_{i,j}$ that
are smaller than the $k$--th smallest element $A_{i,j}(l,m)$, so the
considerations just made, provide the structure of this set. See Figure~\ref{fig:algorithm} for an illustration.

Each partial contour $\tau'$ is thus a sequence of positions in $A_{i,j}$ such that
(i) the rows containing these positions form a contiguous subsequence, starting from the first row
of $A_{i,j}$, (ii) each row in this subsequence has exactly one entry of $\tau'$, and
(iii) the sequence of columns of the entries of $\tau'$ is weakly monotone decreasing:
if $(a,b)$ and $(a+1,b')$ are in $\tau'$ then $b'\le b$.
Any sequence $\tau'$ that satisfies properties (i)--(iii) is called a \emph{valid partial contour}.
Note that a valid partial contour depends only on the positions of the contour in the box $A_{i,j}$, and not on the actual values of the entries of $A_{i,j}$.

Let $\mu'$ be some valid partial contour, as just defined, over a $[g]\times [g]$ position set,
such that the sum of the column indices of positions in $\mu'$ is exactly $k$.
Write $\mu'= (\mu'_{\row} , \mu'_{\col})$, as was done for permutations above, so that $\mu'_{\row}$ gives the row indices of the elements of $\mu'$,
and $\mu'_{\col}$ gives their column indices. Denote by $t'\le g$ the number of positions in $\mu'$.

For given indices $\ell,m$, we can determine, using~(\ref{eq:large}) and~(\ref{eq:small}),
whether $\mu' = \tau'(A_{i,j}(l,m))$, by testing, for each $t\in [t']$, whether
$A_{i,j}(\mu'(t)) < A_{i,j}(l,m)$ and $A_{i,j}(\mu'(t) + (0,1)) > A_{i,j}(l,m)$, except for the $t_0$ for which $A_{i,j}(\mu(t_0)) = A_{i,j}(l,m)$,
since then the second inequality becomes an equality; this takes at most $2t'-2$ comparisons.
By Fredman's trick, and since $A_{i,j}=A_i + A_j$, this can be restated, that $\mu' = \tau'(A_{i,j}(l,m))$ iff the (red) point $p_j$
dominates the (blue) point $q_i$, where
\begin{align}\label{eq:mu}
p_j &= \left(\ldots, \; A_j(m) -A_j(\mu'_{\col}(t)),\;   A_j(\mu'_{\col}(t+ (0,1)) - A_j(m)),    \;\ldots \right)\\
q_i &= \left(\ldots, \; A_i(\mu'_{\row}(t)) - A_i(l),\;\:\:\; A_i(l) - A_i(\mu'_{\row}(t)),   \;\ldots \right) , \nonumber
\end{align}
where the $2t'-2$ coordinates are indexed in pairs by $t\in[t'] - \{t_0\}$.

We regard each box $A_{i,j}$ as being partitioned into $h=g\log g$ sets $A_{(i,j)1}, \ldots, A_{(i,j)h}$, each of size at most $s = g / \log g$, such that
for $ k\in [h]$, $A_{(i,j)k}$ is the set of all elements that
are at least the $(k-1)s$--smallest element, and smaller than the $ks$--smallest element in $A_{i,j}$.
Our goal is to compute, for each box $A_{i,j}$, the positions of the elements of the sets $A_{(i,j)1}, \ldots, A_{(i,j)h}$,
and the correct sorting permutation of each of them,
as well as to determine, for each $k\in [h]$, the position of the $ks$--smallest element in $A_{i,j}$.

Fix $k\in [h]$.
We enumerate all the pairs of realizable valid partial contours $\mu'_{(k-1)s},\, \mu'_{ks}$,
such that (i) $\mu'_{(k-1)s}$ lies to the left and above $\mu'_{ks}$,
and (ii) the sums of the column indices of their entries are $(k-1)s$ and $ks$, respectively.
Let $S_k$ be the set of positions enclosed between the two partial contours
$\mu'_{(k-1)s}$ and $\mu'_{ks}$, excluding $\mu'_{(k-1)s}$ and including $\mu'_{ks}$.
For each $A_{i,j}$, we want to identify the pair $(\mu'_{(k-1)s},\, \mu'_{ks})$,
for which  $\mu'_{(k-1)s}$ and $\mu'_{ks}$ are the partial contours of the $(k-1)s$-smallest and the $ks$-smallest elements of $A_{i,j}$, respectively.
Thus $S_k$ is the set of the $s$ positions of the elements of $A_{i,j}$ that are larger or equal to
the $(k-1)s$-smallest element and smaller than the $ks$-smallest element.
These are the positions of the set $A_{(i,j)k}$. See Figure~\ref{fig:algorithm} for an illustration.

It is easily seen that there are at most $2^{4g}$ pairs of sequences $(\mu'_{(k-1)s},\, \mu'_{ks})$,
and there is only one unique pair of valid partial contours $(\mu'_{(k-1)s}$, $\mu'_{ks})$ that
satisfy all the above requirements for a specific box $A_{i,j}$, as there is only one $(k-1)s$--smallest element and only one
$ks$--smallest element in $A_{i,j}$ (assuming, as above, that all the elements of $A_{i,j}$ are
distinct). We enumerate all pairs of positions $P_1, P_2 \in [g]^2$,
such that $P_1 \in \mu_{(k-1)s}$ and $P_2 \in \mu_{ks}$
(recall that $\mu'$ is a partial contour of some contour $\mu$, where $\mu$ is uniquely determined from $\mu'$, see Figure~\ref{fig:algorithm}).
There are at most $(2g)^2 = 4g^2$ such positions.
We also enumerate every realizable permutation $\pi : [s] \rightarrow S_k$ of
the elements at positions in $S_k$ (where, for each $A_{i,j}$, we want to identify the permutation that sorts its elements at the positions of $S_k$).
The number of permutations is bounded trivially by $s! = (g / \log g)!$.

We now extend the points defined in~(\ref{eq:mu}), to make them encode additional information,
as follows.  For every tuple $(P_1,\, P_2,\, \mu'_{(k-1)s},\, \mu'_{ks},\, \pi)$, we create
red points $\{p_j\}_{j\in [n/g]}$ and blue points $\{q_i\}_{i\in [n/g]}$ in $\mathbb{R}^{4(t'-1) + s-1}$,
such that $p_j$ dominates $q_i$ iff (i) $\mu'_{(k-1)s} = \tau'(A_{i,j}(P_1))$,
(ii) $\mu'_{ks} = \tau'(A_{i,j}(P_2))$, and (iii) $\pi$ is the unique sorting permutation
of the portion of $A_{i,j}$ with indices in $S_k$.
The first $4t'-4$ coordinates
encode the correctness of $\mu'_{(k-1)s}$ and $\mu'_{ks}$ (as in~(\ref{eq:mu}), using the positions $P_1$, $P_2$ as those defining the respective contours),
and the last $s-1$ coordinates encode the correctness of $\pi$,
as in Section~\ref{subsec:GPalgorithm} but for a permutation
of size at most $s = g / \log g$. We do this $h = g\log g$ times, for each $k\in [h]$.

According to Lemma~\ref{lem:redblue}, the overall time to report all bichromatic dominating pairs is
\[
O\left( h\cdot 2^{4g} g^2 s! c^{4(g-1)+s-1}_{\epsilon} (n/g)^{1+\epsilon} + h(n/g)^2 \right).
\]
The second term is the output size, because for each of the $(n/g)^2$ boxes $A_{i,j}$, there will be exactly $h$ dominating pairs, one for each pair of consecutive partial contours, as above.
By fixing $\epsilon = 1/2$ and $g= d\log n$ with a small enough $d$,
the first term will be negligible and the runtime will be dominated by the output size $O(h(n/g)^2) = O(n^2\log g / g) = O(n^2 \log\log n / \log n)$.

We can now search an element $x$ in a box $A_{i,j}$, in $O(\log g)$ time.
We first do a binary search, in $O(\log g)$ time, over the $h$ positions storing the $ks$--smallest element of $A_{i,j}$, for $k\in [h]$ (we have already computed their positions, and, by definition, they are already sorted). This will give us a single set $A_{(i,j)k}$ that can possibly contain $x$.
Then we do another binary search in $A_{(i,j)k}$, also in $O(\log g)$ time, as we already found its sorting permutation earlier.
Note that each such permutation $\pi$ is of length at most $g / \log g$, and of values from $[g]^2$. Thus, by our earlier choice of $g$,
$\pi$ can be stored in a machine word of size $O(\log n)$, and be accessed in $O(1)$ time.
Each element $-A(k)$ is being searched in at most $2\ceil{n/g}$ boxes (as in Steps 4.1--4.3 of Gr\o nlund and Pettie's decision tree, described in Section~\ref{sec:breakthrough}).
Hence, the total running time of the algorithm is $O(n^2\log g /g) = O(n^2 \log\log n / \log n)$ {\em deterministic} time.

\end{document}